\documentclass[letterpaper,referee]{raa}            

\usepackage{graphicx,times}             
\usepackage{natbib}
\usepackage{amssymb,amsmath}

\usepackage[letterpaper]{geometry}
\usepackage[pagebackref=true]{hyperref}
\hypersetup{colorlinks = true, linkcolor = green, anchorcolor = red, citecolor = blue, filecolor = red, urlcolor = red}

\begin{document}

   \title{The Energetics of White-light Flares Observed by SDO/HMI and RHESSI}

   \volnopage{Vol.0 (201x) No.0, 000--000}      
   \setcounter{page}{1}          

   \author{Nengyi Huang
      \inst{1}
   \and Yan Xu
      \inst{1,2}
   \and Haimin Wang
      \inst{1,2}
   }


   \institute{Center for Solar-Terrestrial Research, New Jersey Institute of Technology, Newark, NJ 07102-1982, USA; {\it nh72@njit.edu}\\
        \and
             Big Bear Solar Observatory, Big Bear City, CA 92314, USA\\
   }

   \date{Received~~2015 month day; accepted~~2015~~month day}

\abstract{White-light (WL) flares have been observed and studied more than a century since the first discovery. However, some fundamental physics behind the brilliant emission remains highly controversial. One of the important facts in addressing the flare energetics is the spatial-temporal correlation between the white-light emission and the hard X-ray radiation, presumably suggesting that the energetic electrons are the energy sources.  In this study, we present a statistical analysis of 25 strong flares ($\geq$M5) observed simultaneously by the Helioseismic and Magnetic Imager (HMI) on board the Solar Dynamics Observatory (SDO) and the Reuven Ramaty High Energy Solar Spectroscopic Imager (RHESSI). Among these events, WL emission was detected by SDO/HMI in 13 flares, associated with HXR emission. To quantitatively describe the strength of WL emission, equivalent area (EA) is defined as the integrated contrast enhancement over the entire flaring area. Our results show that the equivalent area is inversely proportional to the HXR power index, indicating that stronger WL emission tends to be associated with larger population of high energy electrons. However, no obvious correlation is found between WL emission and flux of non-thermal electrons at 50 keV. For the other group of 13 flares without detectable WL emission, the HXR spectra are softer (larger power index) than those flares with WL emission, especially for the X-class flares in this group. }


   \authorrunning{N. Huang et al.}            
   \titlerunning{The Energetics of White-light Flares}  

   \maketitle

%
%
\section{Introduction}           

White Light (WL) flares are characterized with sudden emission in the visible continuum against the bright photospheric background.
In the literatures, they are always associated with strong X-ray bursts and therefore are classified in the top group of GOES classification \citep{Neidig1983b}.
However, this empirical statistic is subject to the limitation of observing techniques, for instance the spatial-temporal resolution and dynamic range of detectors.
With the development of both space-based and ground-based instruments, WL flares were becoming more and more resolvable.
\citet{zirin1988} and \citet{Neidig1989} predicted that the WL emission may exist in all the flares at different levels, which were thought to be only detectable in strong flares.
\citet{Matthews2003} surveyed  observations of flares on G-band from Yohkoh and found flare of GOES class down to of C7.8 with brightening signal.
\citet{Hudson2006} investigated flares observed by the Transition Region and Coronal Explorer (TRACE). By removing the UV line contaminations, the authors detected WL emission of a flare in GOES class of C1.6.
\citet{Jess2008} studied C2.0 flare on 2007 August 24 with the high-resolution ground-based Swedish Solar Telescope (SST) and repported that the contrast of WL emission at its peak time above the quiescent flux was as high as 300\%, which appears, however, in a very small area.
\citet{Wang2008} surveyed flares observed by Hinode and suggested that M1 class is the lower limit for the flare with detectable WL emission observed by Hinode.

Since the first discovery of WL flare by Carrington in 1859, tremendous efforts have been made in solving the puzzle of its energetics. It is well accepted that the initial energy release is from the corona by magnetic reconnection, from which huge amount of electrons are accelerated to near relativistic speeds and part of them penetrate to lower atmosphere spiraling along magnetic field lines \citep{Najita1970, Hudson1972}. However, the formation height and mechanisms of the WL emission remain highly controversial. Several models have been proposed to address these questions. For instance, the direct heating model indicates that the continuum emission is generated by the precipitating electrons via collision and ionization in the lower atmosphere \citep{Brown1971}. However, considering the density and short collision range of the upper atmosphere, the electron energy for direct heating should be critically high: 350 KeV electrons can only precipitate to temperature minimum region \citep{Aboudarham1986}. \citet{Emslie1978} calculated the relation between required energies of electrons to penetrate to certain layers and the coresponding densities. As the result, to heat the photosphere ($\tau_{5000} = 1$) the initial energy should be at at least a few MeV. Moreover, to generate a WL flare, the total population of electrons to be accelerated in corona would be too high. Nevertheless the hypothesis of direct heating is not necessarily the only mechanism. A good alternative explanation is that the electrons stop, release their energies and heat the lower layer to power the continuum emission by back warming \citep{Machado1989, Metcalf1990, Ding2003a}.

All the flare models assume that the WL emission is generated by accelerated electrons \citep{Hudson1992, Fletcher2007}.A direct diagnosis of the electron beams is Hard X-ray (HXR) bremsstrahlung radiation. The associations between WL and HXR during flares were proposed by \citet{Rust1975}, and confirmed by observations in both spatial and temporal perspectives \citep[e.g.][]{Neidig1993b,Xu2004a,Hudson2006}. Therefore, coordinated analysis of WL and HXR can reveal the mechanism of how these high-energy electrons should generate bursts of emissions.\citet{Battaglia2011} compared EUV, WL and HXR emission of a limb flare using data from SDO/HMI, SDO/AIA and RHESSI. They found that the EUV emission comes from higher layers (3~Mm from the photoshpere) where low energy electrons ($\sim$ 12 keV) deposit their energies. The HXR of 35-100 keV is suggested in lower layers in a range of 1.7~Mm to 0.8~Mm and the WL locates a little higher at 1.5~Mm. \citet{MartinezOliveros2012} related WL with HXR in a similar energy range of 30-80 keV. However, they found the formation heights were rather lower for both WL and HXR at 195 km and 305 km, respectively. \citet{Cheng2015} further confirms the temporal correlation between WL and HXR emission (26-50 keV) using SDO and {\it Fermi} data, respectively. \citet{Hao2012} studied an M6.3 flare close to the disk center. Their results show a clear Balmer jump and classify this flare as an type I flare \citep{Fang1995a} with direct link to electron precipitation.

Motivated by the above case studies, we attempt to relate the quantitative characteristics of WL and HXR emission \textbf{statistically}. Besides the total energy of carried by accelerated electrons, the distribution of electrons as a function of energy is very important and a key parameter for numerical simulations. By solving the radiative hydrodynamic equations, the atmosphere response to the precipitating electron beams can be simulated. The original idea was developed for stellar flare emission, therefore this kind of simulations are simplified to deal with one dimensional configurations (along the direction of flare loops). The input parameters can be assumed or derived from the HXR observations. Previous studies have matched the simulated results with the observed emission at a certain level \citep[e.g.][]{Allred2005,Cheng2010,RubiodaCosta2016}. Usually, HXR emission composes two components, thermal and non-thermal emission. The non-thermal emission always has power-law spectrum, which can be described as one or more linear functions in a log-log plot. The slope of such a linear function, also known as power index, is the key parameter in determining the contribution of the high energy electrons to the entire HXR emission. To quantitatively denote the WL emission strength, we followed the work of \citet{Wang2008}. For each WL flare, they integrated the contrast over the areas of WL flare kernels and defined this quantity as equivalent area (EA), which describes the strength and extent of WL emissions. In this paper, we present the investigation of a series of WL flares. Comparing and correlating WL and HXR emissions, we try to answer whether direct heating or back warming dominates in powering WL flares.

\section{Observations and Data Reduction}

Solar Dynamics Observatory (SDO; \cite{SDO}) has been monitoring the Sun since its first light in 2010 and flying with three major experiments--Atmospheric Imaging Assembly (AIA), EUV Variability Experiment (EVE), Helioseismic and Magnetic Imager (HMI). Among those three, HMI provides full disk maps of visible intensity and vector magnetograms. The intensity maps are obtained by taking six sampling points across the Fe~{\sc i} absorption line 6173.3~\AA\, estimating the Doppler shift, linewidth, and linedepth, then "reconstructing" the continuum intensity \citep{HMI}. The effective cadence of the visible continuum images is 45 s and the angular resolution is around 1\arcsec. The Reuven Ramaty High Energy Solar Spectroscopic Imager (RHESSI; \cite{RHESSI}) was launched in 2002 to explore hard X-ray (HXR) emissions during solar flares. It is sensitive to an energy range from 3 keV to 17 MeV with various energy resolutions and angular resolutions. The image cadence is usually 4 s, determined from the period of a full cycle of modulation.  In this study, we use SDO/HMI intensity images to retrieve the WL flare signals and RHESSI HXR data for imaging and spectroscopic analysis.

Our targets are flares with WL emission observed by HMI WL channel and have coverage by RHESSI. Although it was predicted theoretically and confirmed observationally that the WL emission is a common feature not only for most violent events, it is still generally accepted that the stronger flares are more likely to be associated with detectable WL enhancement. The reason is that the detection of WL flares depends on the capabilities of instruments, for instance the resolution and dynamic range. \citet{Neidig1983a} found that WL emission can be found only in X-class flares. \citet{Wang2008} extend the lower limit to M-class by studying Hinode observations. Using even large telescope, the 1-m Swedish telescope,  \citet{Jess2008} found strong WL emission in a C-class flare.

According to the resolution conditions of HMI, we set a threshold of M5.0 to select events. In addition, to minimize the projection effect, near limb events ($\geq 50^{\circ}$) are excluded. Also, there are gaps of RHESSI observations, such as satellite night \& South Atlantic Anomaly (SAA). Therefore, we need to eliminate the events which were not covered by RHESSI. In practice, we first select all events covered by RHESSI which occurred close to the disk center. During the period from March 2010 to June 2015, 25 flares satisfied with the criteria including flares from GOES class of M5.1 to X3.1.  Among these events, we observed that the WL signals were present not only the in strongest X-class, but also in some of the M-class flares.  Moreover, some X class flares do not have detectable WL enhancement signal. Therefore, we split the events into two categories, with and without WL emission, as shown in Table~\ref{wl} and Table~\ref{nwl}.

\begin{figure}
 \includegraphics[width=14.0cm, angle=0]{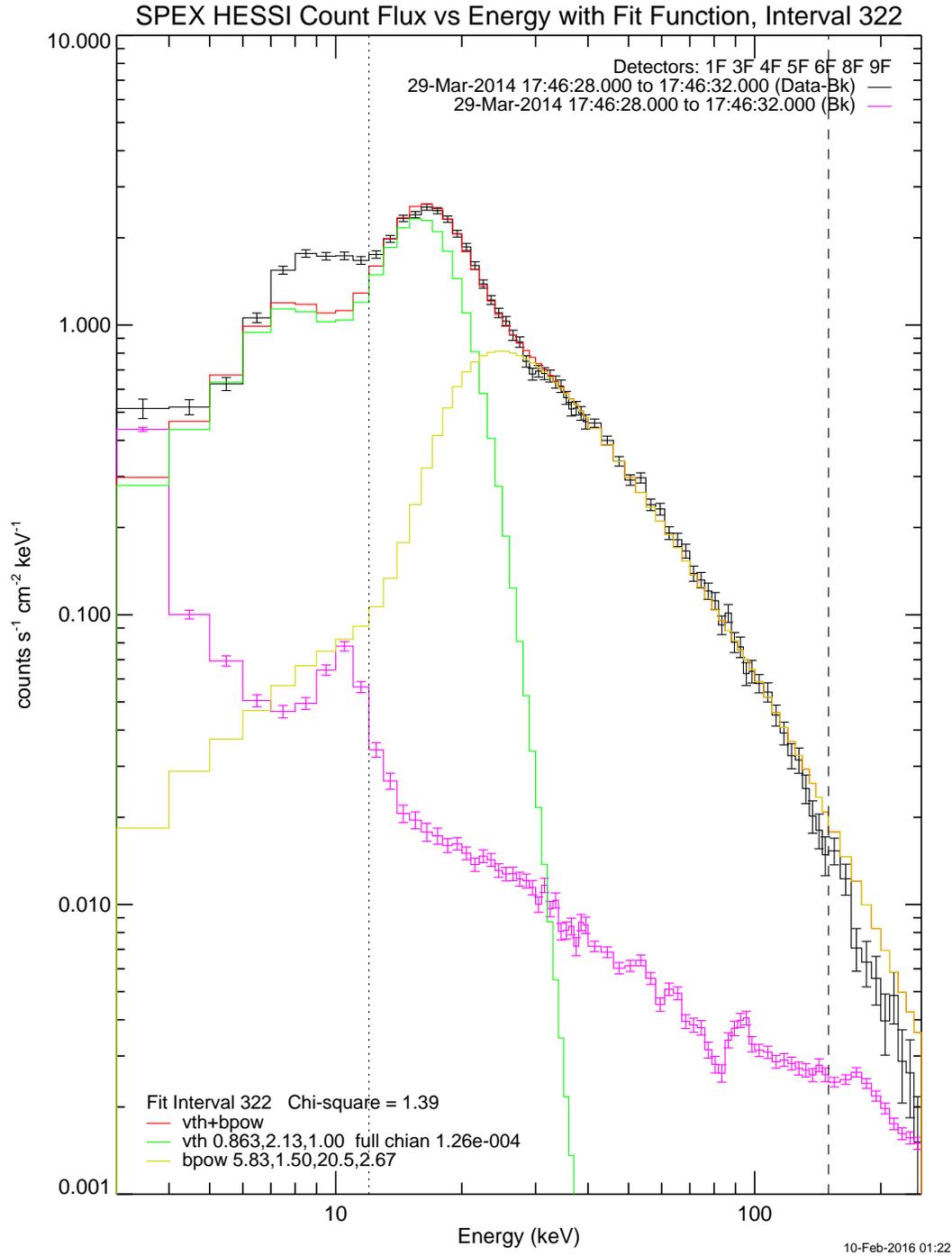}
 \caption{Fitting of peak spectrum of the flare on 2014 March 29th, using two components of Variable Thermal and Non-thermal broken power law. The black curve is the spectrum data after subtracting the background (pink). The modeled thermal, non-thermal components and the overall spectrum are plotted in green, yellow and red colors, respectively.}
 \label{hsi}
\end{figure}

\begin{table}
\begin{center}
\caption{Properties White-Light Flares}
\label{wl}

\begin{tabular}{cllllcc}
\\
 \hline\noalign{\smallskip}
Date       &	Time		&	GOES	&        AR\#	&	Location	&	Equivalent Area		&	HXR Spectral Index\\
 \hline\noalign{\smallskip}

2011.02.15		&		01:43		&		X2.2		&		11158		&		S21W12		&		1.45		&		5.30		\\
2011.07.30		&		02:04		&		M9.3		&		11158		&		N16E32		&		5.55		&		4.84		\\
2011.09.06		&		22:08		&		X2.1		&		11283		&		N16W15		&		6.31		&		2.88		\\
2011.09.08		&		15:32		&		M6.7		&		11283		&		N14W41		&		0.95		&		2.79		\\
2012.07.04		&		09:47		&		M5.3		&		11515		&		S16W15		&		1.42		&		3.45		\\
2012.07.05		&		11:39		&		M6.1		&		11515		&		S17W30		&		1.20		&		2.99		\\
2013.11.10		&		05:07		&		X1.1		&		11890		&		S13W13		&		2.34		&		3.98		\\
2014.01.07		&		10:07		&		M7.2		&		11944		&		S13E13		&		2.52		&		3.65		\\
2014.03.29		&		17:35		&		X1.0		&		12017		&		N10W32		&		3.20		&		2.67		\\
2014.10.22		&		14:02		&		X1.6		&		12192		&		S14E13		&		0.87		&		4.46		\\
2014.10.24		&		21:07		&		X3.1		&		12192		&		S12W31		&		0.71		&		6.25		\\
2015.03.10		&		03:19		&		M5.1		&		12297		&		S15E39		&		0.83		&		2.99		\\
2015.03.11		&		16:11		&		X2.1		&		12297		&		S17E22		&		1.15		&		4.33		\\

 \hline\noalign{\smallskip}\hline
\end{tabular}
\end{center}
\end{table}

\begin{table}
\begin{center}
\caption{Properties of Flares without White-Light emission}
\label{nwl}

\begin{tabular}{cllllc}
\\
 \hline\noalign{\smallskip}
Date       &	Time		&	GOES	&        AR\#	&	Location	&		Spectral Index\\
 \hline\noalign{\smallskip}

2011.02.13		&		17:28		&		M6.6		&		11158		&		S19E01		&		5.05		\\
2011.03.09		&		23:10		&		X1.5		&		11166		&		N10W11		&		5.91		\\
2011.09.25		&		04:31		&		M7.4		&		11302		&		N12E45		&		4.25		\\
2012.03.09		&		3:22			&		M6.3		&		11429		&		N17W01		&		5.17		\\
2012.05.10		&		04:11		&		M5.7		&		11467		&		N10E20		&		3.84		\\
2012.07.12		&		16:28		&		X1.4		&		11520		&		S20W03		&		7.61		\\
2014.01.01		&		18:40		&		M9.9		&		11936		&		S16W45		&		9.40		\\
2014.04.18		&		12:50		&		M7.3		&		12036		&		S15W35		&		2.77		\\
2014.10.22		&		01:16		&		M8.7		&		12192		&		S13E21		&		4.02		\\
2014.10.25		&		16:55		&		X1.0		&		12192		&		S12W31		&		7.73		\\
2014.12.04		&		18:05		&		M6.1		&		12222		&		S20W34		&		5.84		\\
2014.12.18		&		21:58		&		M6.9		&		12242		&		S18W26		&		4.07		\\

 \hline\noalign{\smallskip}\hline
\end{tabular}
\end{center}
\end{table}

\begin{figure}
 \includegraphics[width=14.0cm, angle=0]{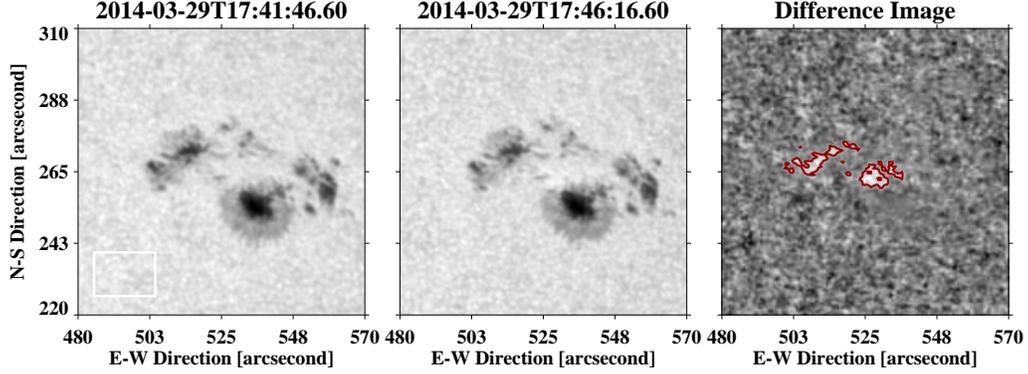}
 \caption{Image processing steps of an example event, X1.0 flare on 2014 March 29th. Left panel: the pre-flare image of the active region, where the region in the white box was selected as the background. Middle panel: the flare peak image taken 5 minutes prior to  the WL flare peak. Right panel: the difference image of the flare peak subtracting the pre-flare image, where the WL flare ribbons were shown in red contours. The total number of flaring pixels is 124, and the equivalent area is 3.20 arcsec$^2$.}
 \label{wlsample}
\end{figure}

For each event, we obtained a full-disk image sequence observed by HMI about 10 minutes prior to the HXR flux peak lasting at least 15 minutes. Then we zoomed in to the flare region with a small FOV including the entire flare sources identified by HXR images. All the images are then aligned spatially using the first frame as a reference. Figure~\ref{wlsample} shows an example of the X1.0 flare on 2014 March 29th. The left panel presents an image taken at 17:41:46 UT, which is before the flare and used as the reference frame for alignment. In order to show the WL emission clearly, difference images are constructed by subtracting each image by the reference frame. This process requires a normalization of the image sequence. To do that, a quiet Sun area was selected, indicated by the white box, and the averaged intensity within this area is defined as the background $I_{b}$. Then the normalized images are obtained and the contrast defined by $Contrast$ = $\frac{I - I_{b}}{I_{b}}$ is retrieved. In the middle panel of Figure~\ref{wlsample}, an image is shown during the peak of the flare. A very weak brightening can be seen around the center but not significant. In the difference image, shown in the right panel, two elongated flare sources appear as outlined by red contours. To quantitatively describe the WL emission, we defined the equivalent area (EA) following \citet{Wang2008}. First, we calculate the standard deviation in the reference area, then the threshold was set as 3 times of standard deviations in this region to obtain the areas of WL sources. Second, we calculate the EA by integrating the contrast enhancement over the entire flaring area (A), $EA$ = $\oint_{A}\,$Contrast~$d(A)$. As an example, the EA for the 2014-03-29 event is 3.20 arcsecond$^2$. The results of all WL events are listed in Table~\ref{wl}.

For events without detectable WL emission, the flare peak times are defined as the maxima in 25-50 keV HXR emission. HXR spectra during the peak times are then constructed using the default setups of RHESSI GUI with pileup correction because all the events are stronger than M5.0. The spectral fitting involves two models, the variable thermal (vth) and the broken power law (bpow), in which the former represents the thermal HXR component usually in low energy ranges, and latter represents the non-thermal component. Our interest is the non-thermal component, representing the high energy part and usually is correlated with HXR footpoint and WL emissions. The representative parameter of the non-thermal component is the power index, which is the absolute value of the fitted line slop in log-log space. In principle, the power index is inversely correlated with the population of high energy electrons. In other words, a higher power index indicates a softer spectrum with relatively less high energy electrons but a lower power index represents a harder spectrum with relatively more high energy electrons. Figure~\ref{hsi} shows a sample spectrum of the 2014-03-29 event. The power index for this event is 2.67. We carried out the spectral analysis and obtained the power indices for all the 25 events and listed the results in Table~\ref{wl}\ \& \ref{nwl}.

\section{Results and Discussion}
\begin{figure}
 \includegraphics[width=14.0cm, angle=0]{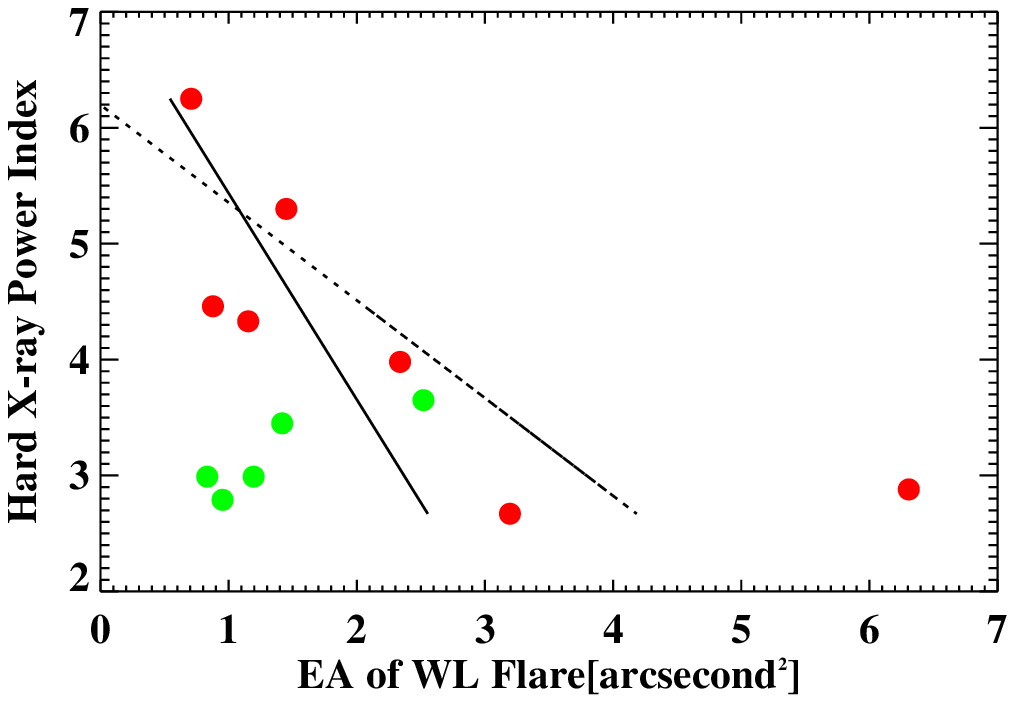}
 \caption{Plot of \emph{equivalent area of WL flare source} vs \emph{hard X-ray power index}. The solid line shows the trend of linear fitting for all events.  A negative correlation is obviously seen (the lower the power index, the larger the EA of WL). The red dots represents the X-class flares and the green dots indicates the M-class flares. The dotted line denotes the trend for X-class flares only.}
 \label{piea}
\end{figure}
\begin{figure}
 \includegraphics[width=14.0cm, angle=0]{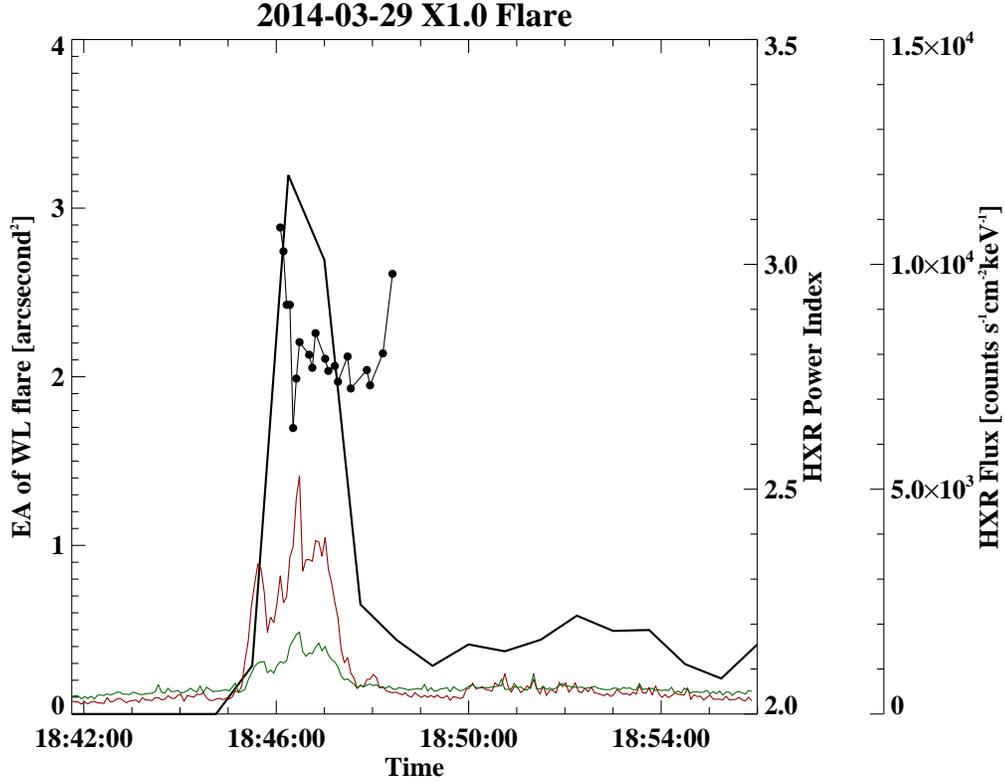}
 \caption{Time evolutions of WL EA, HXR energy flux and HXR power index during the flare on 2014 March 29th. The red and green curves indicate the fluxes of HXR in energy ranges of 50-100 keV and 100-250 keV, respectively. The thick black curve shows the temporal variation of WL emission and the dotted curve represents the power index evolving as a function of time. As expected, a good correlation between the WL and HXR emission is shown. More importantly, we see a negative temporal correlation between WL emission and HXR power index.}
 \label{WL_HXR}
\end{figure}

\begin{figure}
 \includegraphics[width=14.0cm, angle=0]{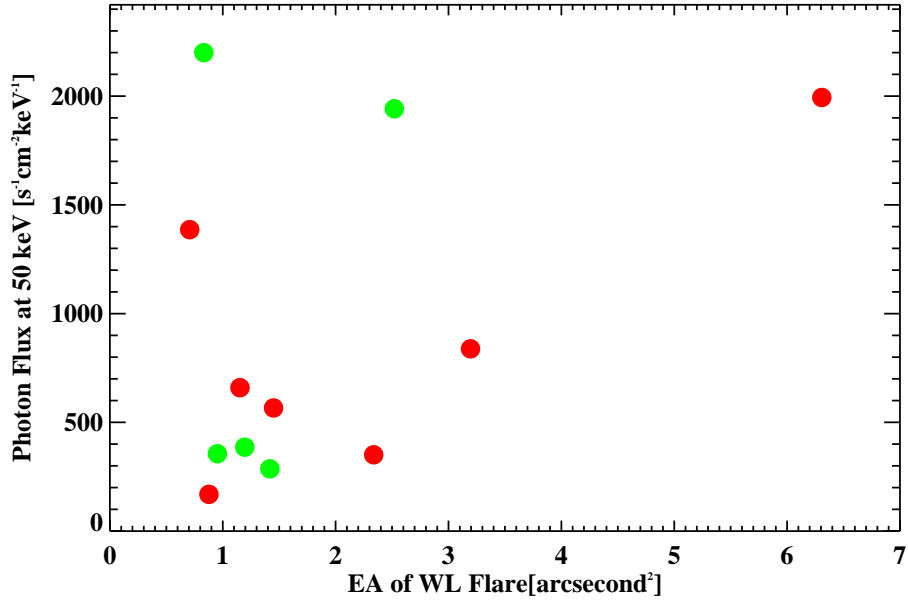}
 \caption{The scatter plot of WL EAs and HXR peak fluxes at 50 keV. The red dots represents the X-class flares and the green dots indicates the M-class flares.}
 \label{fluxes}
\end{figure}
\begin{figure}
 \includegraphics[width=14.0cm, angle=0]{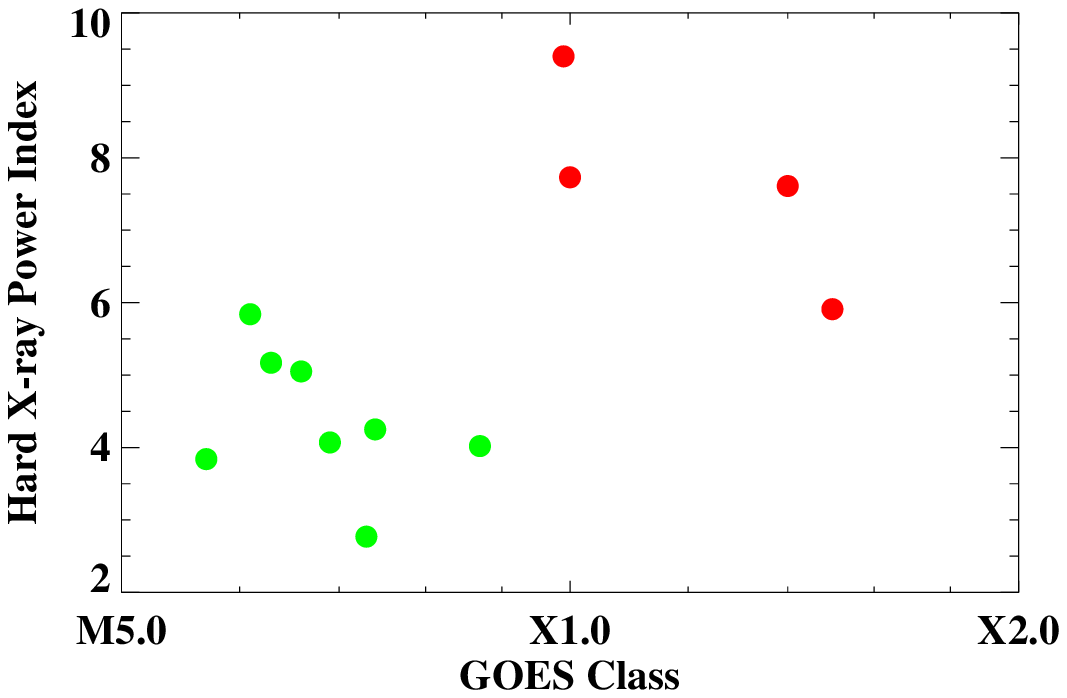}
 \caption{Plot of \emph{GOES class} and \emph{hard X-ray power index} of non-WL flares. For the X-class flares without obvious WL enhancement, the peak spectra show  considerably high HXR power indices, while all of the power indices for M-class flares are smaller than 6.}
 \label{pigoes}
\end{figure}

Among the 25 events we selected, 13 of them are companioned with WL flare signal, and the other 12 are not. Figure~\ref{piea}  shows the scatter plot of WL EA vs. HXR power index.  The red dots represents the X-class flares and the green dots indicates the M-class flares. In principle, we see a correlation between the EA and power index. The solid line denotes the trend for all the events and the dotted line denotes the trend for X-class flares only. According to the trend of X-class flares, the \textbf{negative correlation} between WL strength and HXR power index is more significant. The comparison of EA of WL flares and HXR power index shows that flares with smaller hard X-ray indices tend to have larger equivalent areas. On the other hand, we notice that the HXR spectra are rather hard, with
spectral indices smaller than 4, of all the M-class flares with WL emission shown in this study. We suspect that other M-class flares with relatively higher power index do not have detectable WL emission and are excluded from the analysis. Therefore, this selection effect is one of the reasons why the negative correlation between power index and EA is not obvious for the M-class flares.  As we mentioned in the introduction section, the power index effectively denotes the population of high-energy electrons. Therefore, the negative correlation between equivalent area and power index indicates that the high-energy electrons play a significant role in generating the WL flares, especially for strong (X-class) flares. This result is in favor of the direct-heating model, which requires electrons to have high energy to penetrate down to lower atmosphere and deposit their energy by collision.

In addition to the multiple events analysis, the temporal variations of WL emissions and HXR power indices are retrieved and compared. Figure~\ref{WL_HXR} shows the time evolutions of WL and HXR emissions fluxes and HXR power index during a sample flare on 2014-03-29. The thick solid curve shows that light curve of WL emission, the dotted curve shows the power index evolving as a function of time, and red and green curves plot the time profiles of HXR fluxes in energy ranges of 50-100 keV and 100-250 keV respectively. We see that the WL and HXR fluxes are temporally correlated as expected, and a negative correlation between the WL emission and the power law index, which is typical for all the WL flares in our list.

Furthermore, we looked for the correlation between WL intensities and HXR fluxes for those WL flares. Figure~\ref{fluxes} is the scatter plot of WL EAs and HXR peak fluxes at 50 keV. The red dots represents the X-class flares and the green dots indicates the M-class flares. The correlation between EA  and HXR flux is not as clear based on the events in our study. Our result is consistent with \citet{Fletcher2007}, in which HXR power above 20 keV and 50 keV were compared with the WL power. our result suggests that the distribution of high-energy non-thermal electrons is more important, and the electrons in higher energy are closer related to the generation of WL flares. However, using a larger data set and different method in detecting WL, \citet{Kuhar2016} found a positive correlation between WL emissions and HXR fluxes at 30 keV. Moreover, their result shows no clear correlation between the WL emission and HXR spectral indices, which is also different from our result. The discrepancy is mainly due to the method of measuring the EA of WL emission. For each flare, \citet{Kuhar2016} integrated all the excess WL flux automatically within a certain level of HXR contour. In our study, we manually set thresholds for the WL flare kernels which can be identified visually. Therefore, only the intense flare cores are included in our analysis. According to\citet{Neidig1993a,Xu2006,Isobe2007,Xu2012a}, the WL flare kernels consist bright inner cores and relative weaker halos, corresponding to the direct heating by electrons and the back-warming emission, respectively. The intensity of halo structure is not a monotonically-increasing function of electron energy. Therefore, the relationship between HXR spectral index and WL intensity is altered for halos. By concentrating on the core emission, we exclude the contamination of halos and other uncertainties introduced by the misalignment among frames in a time sequence.

The HXR power indices and the WL EA of the other 12 non-WL flares are plotted in Figure~\ref{pigoes}. Again, the green and red colors are used for M-class and X-class flares, respectively. It is clear that two distinct groups can be defined for M- and X-class flares. For the X-class flares without detectable WL enhancement, we see that all of the four events have relatively higher power indices (\textgreater 6), which implies that they have less high-energy electrons. This results is consistent with the result of WL flares discussed above. However, for M-class non-WL flares, their power indices can be small indicating the spectra are as hard as the WL flares.

In summary, we surveyed disk-center events above GOES class of M5.0 after SDO was launched and covered by RHESSI, and 25 were found. We calculated the EA to quantitatively describe the strength of WL emission, as well as the power index of HXR to denote the population of high-energy non-thermal electrons which have higher penetrating capability. Comparing the EA and power index, we found a negative correlation between them. This result suggests that the high-energy electrons play important role in the flare heating, especially in producing WL emission of strong flares. On the other hand, our result of the M-class flare, especially those without WL emission, confirms that the power index is not the only or most important parameter to determine the WL emission. Considering the complexity of WL and HXR emission, our results in this study are preliminary and can be improved by carrying out more comprehensive analysis. For instance, the $hmi.Ic\_45s$ data from HMI/SDO is not real continuum data but reconstructed by intensity data at six spectral points, which can introduce uncertainties of intensity measurements. For the future analysis, more precise results can be obtained by including filtergrams without any reconstruction and using higher spatial-temporal resolutions.

\begin{acknowledgements}

We would like to thank the referee for valuable comments and suggestions. Obtaining the excellent data would not have been possible without the help of the SDO and RHESSI teams.This work is supported by NSF grants AGS-1539791, AGS-1250374, AGS-1408703 and AGS-1348513.

\end{acknowledgements}

\bibliographystyle{raa}
\bibliography{reference}
\clearpage


\end{document}